# Regarding how Tycho Brahe noted the Absurdity of the Copernican Theory regarding the Bigness of Stars, while the Copernicans appealed to God to answer that Absurdity

Christopher M. Graney
Jefferson Community & Technical College
1000 Community College Drive
Louisville, KY 40272 (USA)
christopher.graney@kctcs.edu

Tycho Brahe, the most prominent and accomplished astronomer of his era, made measurements of the apparent sizes of the Sun, Moon, stars, and planets. From these he showed that within a geocentric cosmos these bodies were of comparable sizes, with the Sun being the largest body and the Moon the smallest. He further showed that within a heliocentric cosmos, the stars had to be absurdly large — with the smallest star dwarfing even the Sun. (The results of Tycho's calculations are illustrated in this paper.) Various Copernicans responded to this issue of observation and geometry by appealing to the power of God: They argued that giant stars were not absurd because even such giant objects were nothing compared to an infinite God, and that in fact the Copernican stars pointed out the power of God to humankind. Tycho rejected this argument.

*The stars, to the naked eye, present diameters varying from a quarter of a minute of space, or less, to as much as two minutes. The telescope was not then invented which shows that this is an optical delusion, and that they are points of immeasurably small diameter. It was certain to Tycho Brahé, that if the earth did move, the whole motion of the earth in its orbit did not alter the place of the stars by two minutes, and that consequently they must be so distant, that to have two minutes of apparent diameter, they must be spheres as great a radius at least as the distance from the sun to the earth. This latter distance Tycho Brahé supposed to be 1150 times the semi-diameter of the earth, and the sun about 180 times as great[*] as the earth. Both suppositions are grossly incorrect; but they were common ground, being nearly those of Ptolemy and Copernicus. It followed then, for any thing a real* Copernican *could show to the contrary, that some of the fixed stars must be 1520 millions of times as great as the earth, or nine millions of times as great as they supposed the sun to be.… Delambre, who comments with brief contempt upon the several arguments of Tycho Brahé, has here only to say, 'We should now answer that no star has an apparent diameter of a second.' Undoubtedly, but what would you have answered* then*, is the reply. The stars were spheres of visible magnitude, and are so still; nobody can deny it who looks at the heavens without a telescope; did Tycho reason wrong because he did not know a fact which could only be known by an instrument invented after his death?*

> — from the article "Tycho Brahé" in the 1836 *Penny Cyclopædia of the Society for the Diffusion of Useful Knowledge* (page 326)

---

[*] in terms of volume

There was a big problem with the Copernican theory.  If, like Nicolaus Copernicus (1473-1543) said, the Earth moves annually about the Sun, and the Sun and the stars are fixed in place, then Earth's annual motion should reveal itself in the stars — a phenomenon known as *annual parallax*.  As orbital motion carries Earth toward certain stars and away from others, those stars should grow brighter and fainter, respectively.  The positions of stars should be different when viewed from one side of Earth's orbit than when viewed from the other.  Indeed, Copernicus said that the variations in brightness and motion of the planets were manifestations of just such effects.  And yet, said Copernicus,

> None of these phenomena appears in the fixed stars.  This proves their immense height, which makes even the sphere of the annual motion, or its reflection, vanish from before our eyes.  For, every visible object has some measure of distance beyond which it is no longer seen, as is demonstrated in optics.  From Saturn, the highest of the planets, to the sphere of the fixed stars there is an additional gap of the largest size [Copernicus 1543, 133].

Yet more than just the Saturn-to-stars gap was of "the largest size" in the Copernican theory; as Tycho Brahe (1546-1601) pointed out, the stars themselves had to be of the largest size.  This was the big problem with the Copernican theory — a problem of "bigness" itself.  To solve the problem of bigness, Copernicans invoked God — God's power was plenty big to create big stars, and the bigness of the stars proclaimed the still bigger bigness of God.  This solution did not impress anti-Copernicans such as Tycho and Giovanni Battista Riccioli.

In the time of Tycho and Copernicus, and indeed throughout most of the history of astronomy, those who took time to look carefully at the stars saw bodies of varying size, or "magnitude" or "bigness".  There were prominent stars such as Arcturus or Vega, and less prominent ones such as Polaris or Albireo, and ones barely visible to the eye, such as Alcor.  Stars like Vega and Arcturus were the top-tier, first-class stars — stars of the first magnitude.  Lesser stars were classified as being of the second, third, fourth, and fifth tiers, or magnitudes.  Those barely visible to the eye were of the sixth magnitude.  This system dated to ancient times.  The lines of demarcation between magnitude ranks were not clear, but what was clear was that stars such as Vega were larger than stars such as Albireo which were in turn larger than stars such as Alcor, for that is how they appear.  References to the magnitudes of stars from centuries ago speak in terms of size or bigness of stars more than in terms of their brightness, whether those references are simply explaining the concept of magnitude —

> The *fixed Stars* appear to be of different Bignesses, not because they really are so, but because they are not all equally distant from us. Those that are nearest will excel in

> Lustre and Bigness; the more remote *Stars* will give a fainter Light, and appear smaller to the Eye [Keill 1739, 47].

— or providing a description of the changing light output of a nova —

> After [the nova] had thus absolutely disappeared, the place, where it had been seen, continued six months vacant. On the seventeenth of March following, the same observer saw it again, in exactly the same place, equal to a star of the fourth magnitude. On the third of April, 1671, the elder Cassini saw it, it was then of the bigness of a star of the third magnitude, he judged it to be a little less than [the star] in the back of the constellation; but, on the next day, repeating the observation, it appeared to him very nearly as large as that, and altogether as bright; on the ninth it was somewhat less; on the twelfth it was yet smaller, it was then less than the two stars at the bottom of Lyra; but, on the fifteenth, it had increased again in bigness, and was equal to those stars; from the sixteenth to the twenty-seventh of the same month he observed it with a peculiar attention; during that period it changed bigness several times, it was sometimes larger than the biggest of those two stars, sometimes smaller than the least of them, and sometimes of a middle size between them. On the twenty-eighth of the same month it was become as large as the star in the beak of the Swan, and it appeared larger from the thirtieth of April to the sixth of May. On the fifteenth it was grown smaller; on the sixteenth it was of a middle size between the two, and from this time it continually diminished till the seventeenth of August, when it was scarce visible to the naked eye ["New Stars" in Hill 1775].

And herein lies the Copernican theory's big problem. Suppose a prominent star such as Aldebaran has bigness such that twenty Aldebarans, arranged side by side, would equal the diameter of the Moon, as seen in Figure 1. As the Moon and Sun are approximately equal in apparent diameter, the apparent diameter of Aldebaran is approximately a twentieth that of the Sun. Supposing Aldebaran and the Sun to be similar sorts of bodies, of similar physical size, the distance to Aldebaran would be twenty solar distances (twenty *Astronomical Units*, or 20 *A.U.*). But were Aldebaran more distant, geometry would dictate its physical size to necessarily be larger: A distance to Aldebaran of 40 A.U. would mean an Aldebaran twice the Sun's diameter; a distance of 100 A.U. would mean a physical size for Aldebaran of five solar diameters. Copernicus's fixed stars at "immense height" solved the annual parallax problem, but at the cost of stars of physical size far in excess of even the Sun.

Tycho Brahe made this cost ever so clear. Tycho was the leading astronomer of his era. While other well-known astronomers like Copernicus or Galileo made their observations and published their results and ideas as individuals, Tycho ran a major observatory ("Uraniborg") and research program on his island of Hven, whose cost to the Danish crown was proportionately comparable to the budget of NASA (Couper, Henbest, and Clarke 2007, 120; Thoren 1990, 188). With access to the biggest and best available instruments, and with the most skilled assistants, Tycho could achieve incredible accuracy in his work: Modern analysis of his work shows that he could define any star's position within a circle of diameter less than a minute of arc (1/60 degree; 1/30 the apparent diameter of the Moon); for certain sorts of measurements, such as the altitude of the North Celestial Pole, he could exceed this accuracy by better than an order of magnitude (Maeyama 2002, 118-9). Owen Gingerich often illustrates Tycho's incomparable contribution to the astronomy of his time by means the 1666 *Historia Coelestis* of Albertus Curtius, which as Gingerich notes contains a few tens of pages of pre-Brahean observations, hundreds of pages of Tychonic material, and again a few tens of pages of post-Brahean observations to 1630. He argues that "only twice in the history of astronomy has there been such an enormous flood of new data that just changed the scenes" — the flood from Tycho Brahe and the flood from the today's digital revolution (Gingerich 2009, 10:00 mark and following). Gingerich has noted that Tycho's quest for better observational accuracy "places him far more securely in the mainstream of modern astronomy than Copernicus himself [Gingerich 1973, 87]".

Tycho obtained precise measurements of the apparent diameters of the fixed stars, determining that a typical first-magnitude star has an apparent diameter of two minutes of arc — one fifteenth the diameter of the Moon or Sun. In a geocentric universe, fixed stars could lie just beyond Saturn (Figure 2) — a distance of just over 12.5 A.U. Thus Tycho determined that the physical diameter of the typical first-magnitude star was about 80% that of the Sun — one of the larger bodies in a celestial assemblage whose smallest member was the Moon and whose largest was the Sun (see Table 1). But in a Copernican universe, in order for annual parallax to be no more than a minute of arc (just falling under Tycho's circle of general accuracy, and thus just evading detection), the distance to the fixed stars would have to be almost 7,000 A.U. Copernicus's Saturn-to-stars "gap of the largest size" would be over 700 times the Sun-to-Saturn distance. And the stars themselves, rather than falling within the size range of the other heavenly bodies, would have to be hundreds of times the diameter of the Sun (see Table 2). What's more, said Tycho, what if the parallax turns out to be smaller than that minute of arc? Then the fixed stars would have to be still larger. Such immense stars at such immense distances were absurd (see Blair 1990, 364; Moesgaard 1972, 51; Brahe 1601, 167).

But according to Christoph Rothmann (~1555-~1600), the German Copernican against whom Tycho leveled this argument, this was not absurd at all. The Creator need not make Creation conform to our notions of reasonableness (Moesgaard 1972, 52). Said Rothmann —

But as far as I am concerned … why should it seem untrue for the distance from the Sun to Saturn to be contained so many times between Saturn and the remoteness of the fixed Stars? or what is so absurd about a Star of the third Magnitude having size equal to the whole annual orb?  What of this is contrary to divine will, or is impossible by divine Nature, or is inadmissible by infinite Nature?  These things must be entirely demonstrated by you, if you will wish to infer from here anything of the absurd.  These things which common men see as absurd at first glance are not easily charged with absurdity, for in fact divine Sapience and Majesty is far greater than they understand.  Grant the Vastness of the Universe and the Sizes of the stars to be as great as you like — these will still bear no proportion to the infinite Creator.  It reckons that the greater the King, so much more greater and larger the palace befitting his Majesty.  So how great a palace do you reckon is fitting to GOD [Brahe 1601, 186; Graney 2012]?

Rothmann was not the first Copernican to invoke "palace of God" imagery in regards to the enormous stars demanded by the Copernican theory.  Thomas Digges (1546-1595) of England — one of only perhaps fifteen identifiable Copernicans in Tycho's time, one of even fewer to write publicly on the theory, and the first to write on it in a vernacular language (Danielson 2006, 232; Wernham 1968, 461) — described the stars in supernatural terms (see Figure 3).  Indeed, Copernicus himself had spoken of the stars in such terms:  "So vast, without any question, is the divine handiwork of the most excellent Almighty [Copernicus 1543, 133]."

However, Tycho was most unreceptive to the use of God to solve the problem of the bigness of stars.  He asks where in nature — where all things are well-ordered in all ways of time, measure, and weight, and there is nothing empty, nothing irrational, nothing disproportionate or inharmonious — do we see the Will of God acting in an irregular or disorderly manner?  It is true, Tycho says, that a finite world can bear no proportion to an infinite Creator, but nature does show proportion and symmetry within itself — and as an example he cites the human body illustrated in the work of the artist Albrecht Dürer.  There is nothing proportional or harmonious or rational, says Tycho, in the Copernican theory's so distant stars that so dwarf the Sun (Brahe 1601, 191-2).

Despite Tycho's exhortations, Copernicans continued to connect the bigness of stars to the power of God.  Several decades after the German Copernican Rothmann spoke of gigantic stars using the language of the "palace of God", and well after the advent of the telescope (Figure 4), the Dutch Copernican Philips Lansbergen (1561–1632) could be found using the same language in his 1629 *Considerations on the Diurnal and Annual Rotation of the Earth, as well as on the True Image of the Visible Heaven; Wherein the Wonderful Works of God are Displayed*.  In this widely read and influential book (the first in Europe whose purpose was popularizing the Copernican theory among a non-mathematical audience), Lansbergen accepts the immense sizes of the stars, as to him these show the divine nature of the heavens. He determines the heavens to be threefold, owing to a reference in 2 Corinthians 12:2 to a "third heaven". The first heaven, says Lansbergen, is that of the planets. The second is that of the fixed stars. It is immense compared to the planetary heaven; each star is indeed the size of Earth's orbit (as Tycho had said must be the case if Copernicus was right). The light of those stars illuminates the whole of the second heaven, which is therefore full of immense splendor. The purpose of this immense size and splendor is to indicate God's infinity to humankind. The heavens, Lansbergen says, echoing the words of Digges and Rothmann before him, are like a fore-court in front of God's palace. The third heaven, that of God, is to the second heaven of the stars as that second heaven is to the first heaven of the planets (Vermij 2007, 124-5).

Thus when Giovanni Battista Riccioli (1598-1671) in his 1651 *Almagestum Novum* reprised Tycho's argument on the bigness of the fixed stars — now using precise telescopic measurements of their diameters and maximum annual parallax (see Figure 4), but obtaining essentially the same result: that in a geocentric cosmos the sizes of stars were consistent with the Sun, Earth, and planets, while in a heliocentric cosmos they dwarfed the Sun (Graney 2010b) — he also reprised Tycho's complaint about how Copernicans answered the star bigness problem. Since nothing is beyond the power of God the Copernican answer was beyond refute in one sense, but, like Tycho, Riccioli rejected that answer to the star bigness problem, stating that "even if this falsehood cannot be refuted, nevertheless it cannot satisfy the more prudent men" (Graney 2012).

Eventually of course the bigness problem would be solved. Shortly after Riccioli's *Almagestum Novum* appeared in print, Jeremiah Horrocks' observations of the moon passing through the Pleiades were published. Horrocks noted that stars being occulted by the moon disappeared not gradually as their apparent bigness required, but all at once. After this Riccioli may have weakened in his enthusiasm for Tycho's argument (Graney 2010b, 462-3, note 39, etc.), and by the early 18$^{th}$ century some astronomers had definitely accepted that the bigness of stars was, in the words of Edmund Halley, an "Optick Fallacy" (Graney & Grayson 2011, 354-356).

Today we understand the distances to stars to be so great that stars are essentially point sources of light, whose apparent size is indeed an illusion born of optics. We measure those vast distances, which span hundreds of thousands of A.U. and more, by means of annual parallax, which was first detected in the 19$^{th}$ century. We know the Sun to be a middling star — there are stars both larger and smaller than it — and were it at stellar distances it too would be essentially a point source of light. We think of the magnitude system as an odd system for measuring the brightness of those points of light, a system in which the brighter the star the lower the magnitude. We say that Tycho's opposition to the Copernican theory was related to religion or the world view of his time or just his inability to take the next logical step.[†] We do not even think of the bigness of stars, and how that bigness once rendered the heliocentric theory so irrational, inharmonious, and disproportionate that Copernicans could only appeal to the power of God to explain it all.

ACKNOWLEDGEMENT: I thank Christina Graney for her invaluable assistance in translating the works of Tycho, Rothmann, and Riccioli from Latin into English.

---

[†] For example, the web page of the Tycho Brahe Museum (located at the site of Tycho's observatory on Hven) has a page on "Tycho Brahe's Worldview", which speaks of Tycho's geocentrism in terms of the established world order, mentions Tycho saying the Bible had as much authority as science, and so forth ("Tycho Brahe´s verdensbillede", 2011). It is not difficult to find statements about how Brahe "could not bring himself" to accept the Copernican theory, both in the popular press (for example, Wertheim 2003), and in scholarly works (for example, Freedberg 2002, 83).

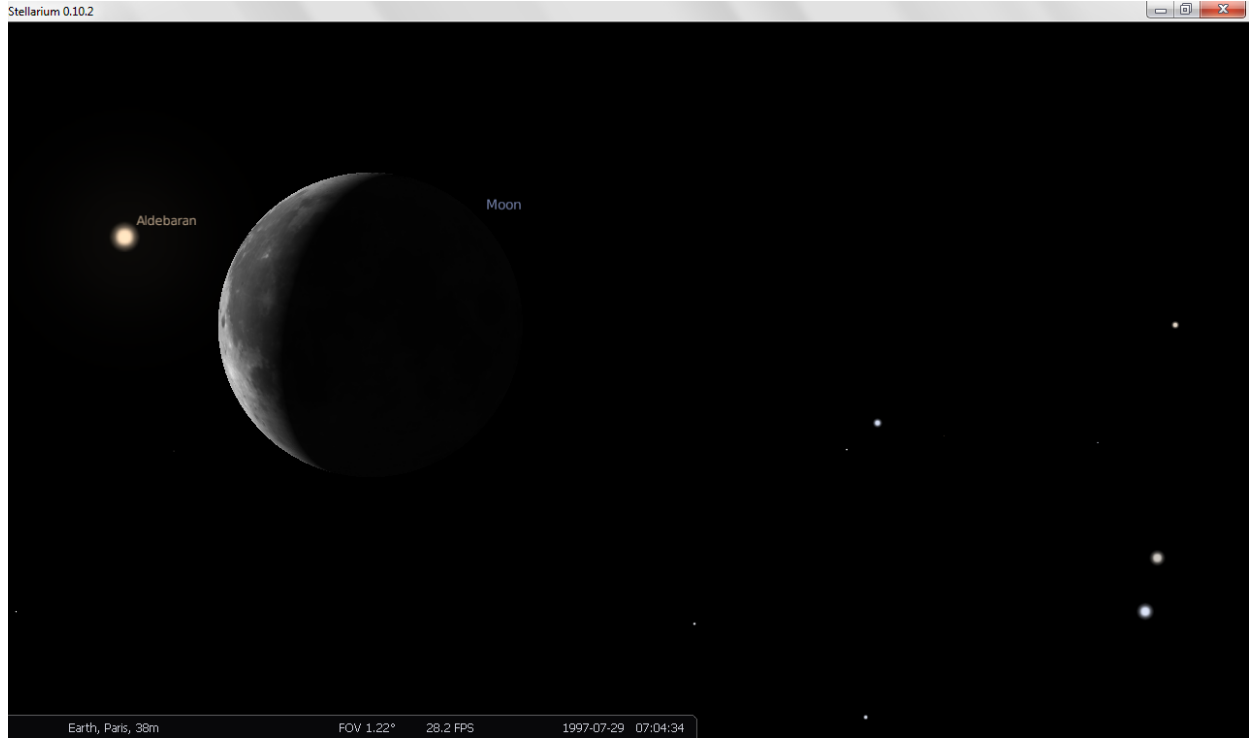

**FIGURE 1**

**Above: The Moon near Aldebaran as represented by the *Stellarium* planetarium software. The diameter of Aldebaran in this representation is about one twentieth the diameter of the Moon; the lesser stars to the right of the Moon are represented with smaller diameters. Stellarium purports to show "a realistic sky … just like what you see with the naked eye" ("Stellarium", 2011).**

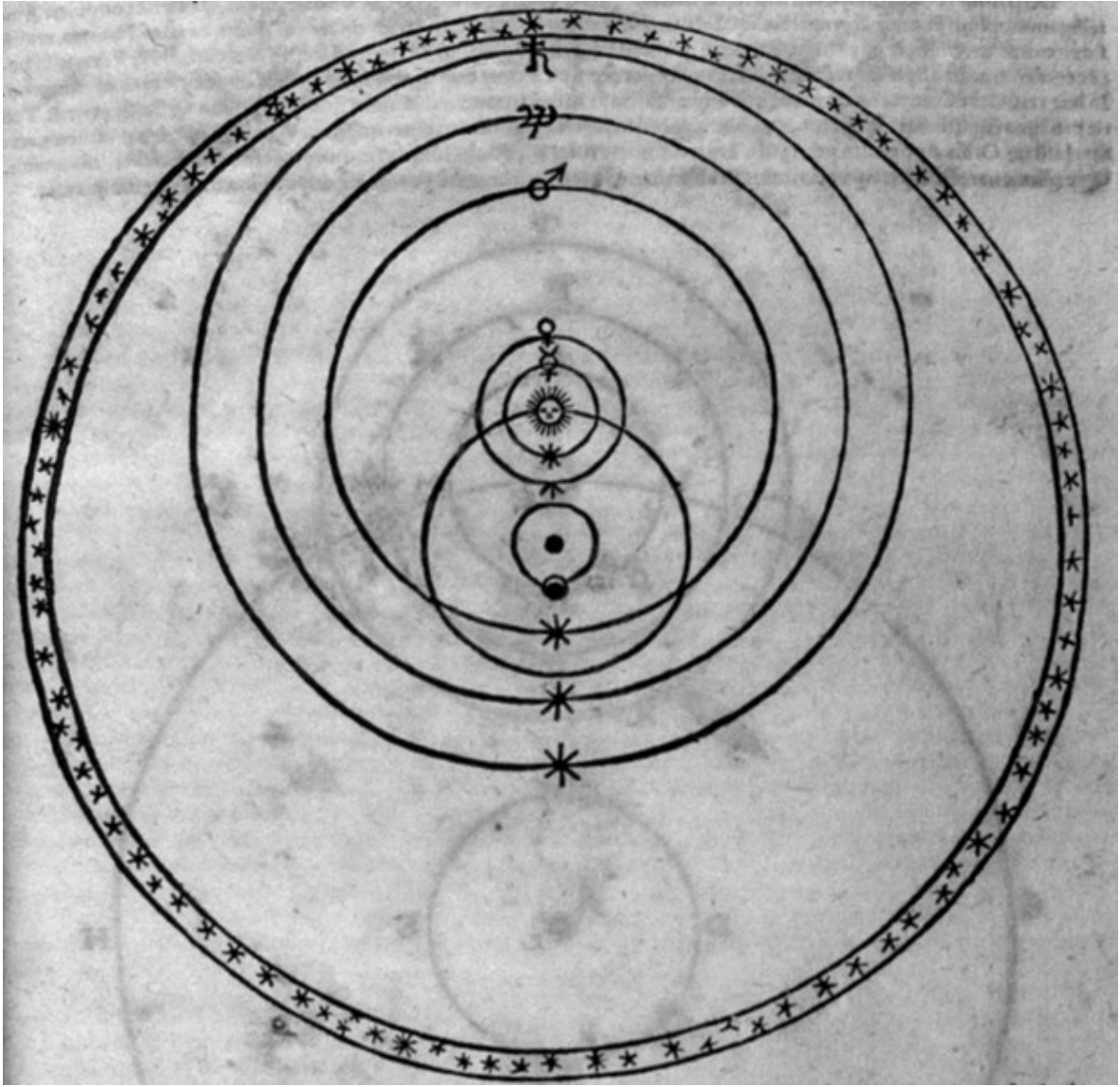

**FIGURE 2**

**The Tychonic geocentric theory. The Moon, Sun, and fixed stars circle the Earth; the planets circle the Sun. As regards the Moon, Sun, and planets, this theory is mathematically and observationally identical to the Copernican theory. Thus it would be fully compatible with Galileo's telescopic observations, such as his observations of the phases of Venus that showed Venus to circle the Sun, or his observations of the moons of Jupiter which showed that heavenly bodies do not all circle the Earth.**

|           | Apparent Diameter | | Distance | | Physical Radius | Physical Volume |
|-----------|------|------|-------|-------|------|--------|
|           | min  | sec  | E.R.  | A.U.  | E.R. | E.V.   |
| Moon      | 33   |      | 60    | 0.05  | 0.29 | 0.02   |
| Sun       | 31   |      | 1150  | 1.00  | 5.19 | 139.40 |
|           |      |      |       |       |      |        |
| Mercury   | 2    | 10   | 1150  | 1.00  | 0.36 | 0.05   |
| Venus     | 3    | 15   | 1150  | 1.00  | 0.54 | 0.16   |
| Mars      | 1    | 40   | 1745  | 1.52  | 0.42 | 0.08   |
| Jupiter   | 2    | 45   | 5990  | 5.21  | 2.40 | 13.75  |
| Saturn    | 1    | 50   | 10550 | 9.17  | 2.81 | 22.26  |
|           |      |      |       |       |      |        |
| 1st mag   | 2    |      | 14400 | 12.52 | 4.19 | 73.50  |
| 2nd mag   | 1    | 30   | 14400 | 12.52 | 3.14 | 31.01  |
| 3rd mag   | 1    | 5    | 14400 | 12.52 | 2.27 | 11.68  |
| 4th mag   |      | 45   | 14400 | 12.52 | 1.57 | 3.88   |
| 5th mag   |      | 30   | 14400 | 12.52 | 1.05 | 1.15   |
| 6th mag   |      | 20   | 14400 | 12.52 | 0.70 | 0.34   |

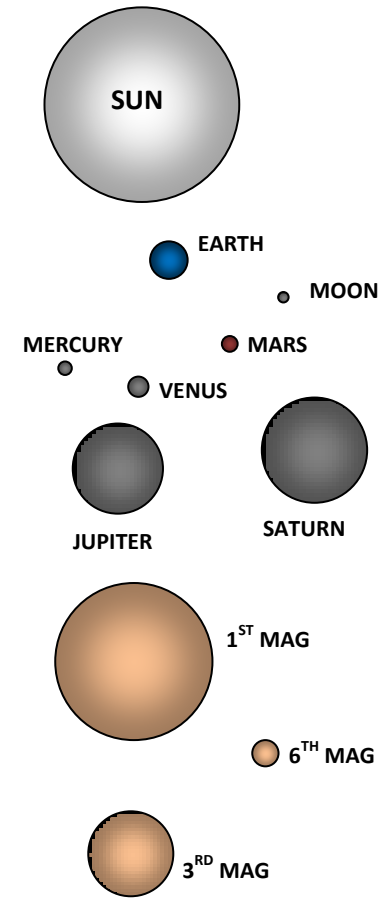

**TABLE 1**

**Tycho Brahe's apparent sizes of and average distances to celestial bodies, with resulting physical sizes assuming a geocentric cosmos (E.R. — Earth Radius; A.U. — Astronomical Unit/solar distance; E.V. — Earth Volume). Note that Tycho reports a first magnitude star to have an apparent diameter approximately one fifteenth that of the Moon. In regard to physical size, the Sun is the largest body; the Moon is the smallest; star sizes fall reasonably within this range, as illustrated by the figures to the right of the table. Tycho reported physical sizes in terms of both radius and volume. Apparent diameter and distance values are from Brahe 1602, 424-431; also see Dreyer 1890, 190-1, and Thoren 1990, 302-7. Physical radius and volume are calculated from these, and so may differ slightly from Brahe's figures, which suffer from rounding and typographical errors.**

      Each representative planetary distance is the average of each planet's two extremes from Earth. For Mercury and Venus, this is simply 1 A.U. For Mars, Jupiter, and Saturn, this equals the radius of their circles of motion around the Sun. Planetary distances in A.U. could be worked out by observing planetary motions — measuring the maximum angle between Mercury and the Sun yields the radius of Mercury's circle, for example — and the values here generally agree with modern values. However, relating A.U. to E.R. was problematic. All methods of determining this value which might work in theory (such as determining the lunar distance in E.R. via triangulation from different observing locations on Earth, exactly measuring the angle between the Moon and Sun when the Moon is precisely at first quarter phase, and then determining the ratio of solar to lunar distances based on how much less that angle is than 90°) were highly prone to error in practice. Tycho used a value of 1150 E.R. = 1 A.U., which is too small by more than a factor of twenty, and leads to a physical size for

the Sun which is likewise too small.  However, Tycho's value was in line with the values used by astronomers from Ptolemy in ancient times to Copernicus; Galileo would use a similar number as well.  See Thoren 1990, 302-4.

In contrast to the other distances, Tycho's stellar distances are not based on measurement at all.  In a geocentric cosmos there is virtually no means of determining stellar distances from Earth.  Geocentrists assumed the fixed stars lay beyond the furthest retreat of Saturn, but that could be 11 A.U. as easily as 12.52 A.U. (Thoren 1990, 304-6) — and were that the case the physical sizes of the stars would be slightly smaller.  Thus the physical star sizes given here are necessarily estimates, based on an assumed interval between Saturn and the fixed stars.

|  | Apparent Diameter | | Distance | | Physical Radius | | Physical Volume |
|---|---|---|---|---|---|---|---|
|  | min | sec | E.R. | A.U. | E.R. | A.U. | E.V. |
| 1st mag | 2 |  | 7,906,818 | 6,875 | 2,300 | 2.00 | 12,167,000,000 |
| 2nd mag | 1 | 30 | 7,906,818 | 6,875 | 1,725 | 1.50 | 5,132,953,125 |
| 3rd mag | 1 | 5 | 7,906,818 | 6,875 | 1,246 | 1.08 | 1,933,658,782 |
| 4th mag |  | 45 | 7,906,818 | 6,875 | 863 | 0.75 | 641,619,141 |
| 5th mag |  | 30 | 7,906,818 | 6,875 | 575 | 0.50 | 190,109,375 |
| 6th mag |  | 20 | 7,906,818 | 6,875 | 383 | 0.33 | 56,328,704 |

**TABLE 2**

**Tycho Brahe's apparent sizes of and average distances to the fixed stars, with resulting physical sizes assuming a heliocentric cosmos (E.R. — Earth Radius; A.U. — Astronomical Unit/solar distance; E.V. — Earth Volume). Compare to Table 1. As the Copernican heliocentric and Tychonic geocentric theories were mathematically identical regarding the Sun, Moon, and planets, values for those bodies are the same in both. Thus in a Copernican cosmos the physical sizes of the stars dwarf even the Sun, as illustrated by the figures below and on the following page. As Tycho pointed out, a third magnitude star must be comparable to the orbit of Earth: 240 times the diameter of the Sun. Apparent diameter and distance values are from Brahe 1602, 424-431; also see Dreyer 1890, 190-1, and Thoren 1990, 302-7. Physical radius and volume are calculated from these, and so may differ slightly from Brahe's figures, which suffer from rounding and typographical errors.**

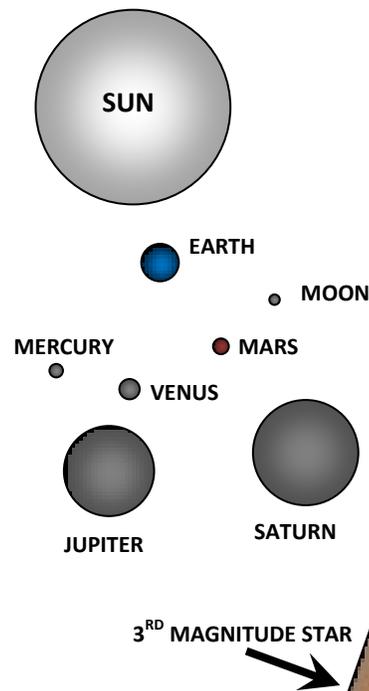

     In a heliocentric hypothesis, the distance to a star can theoretically be determined via the star's annual parallax. Even if parallax is not observed, its non-detection can be used to establish a minimum stellar distance. The stellar distance given here is calculated based on the assumptions that parallax amounts to one minute of arc and that stars lie on a sphere as Copernicus said (and thus are all equidistant from the Sun). Were the parallax actually less than a minute, or were the stars at varying distances as Thomas Digges supposed (Figure 3), then the physical sizes of the stars would be still larger.

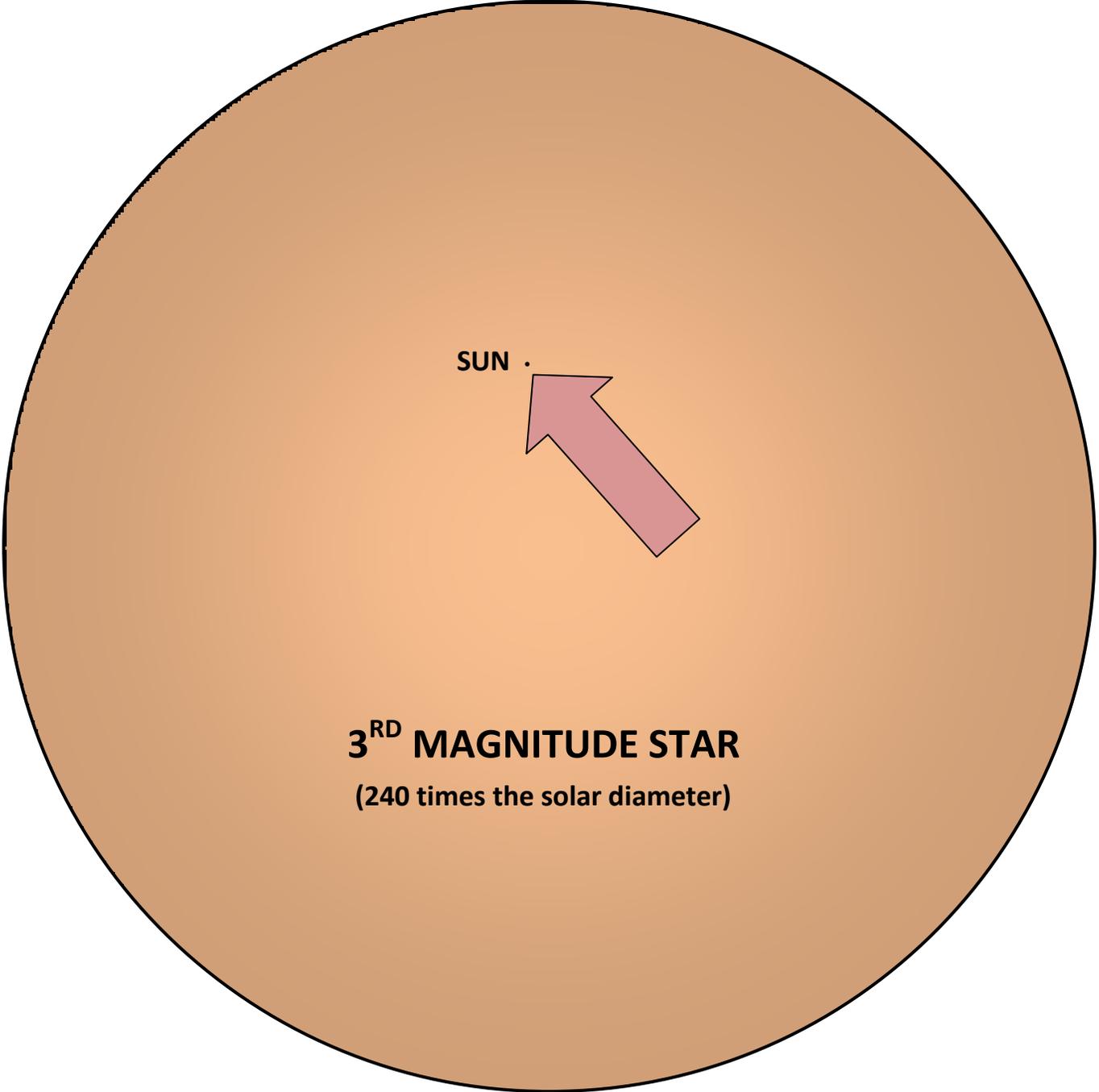

SUN

3<sup>RD</sup> MAGNITUDE STAR

(240 times the solar diameter)

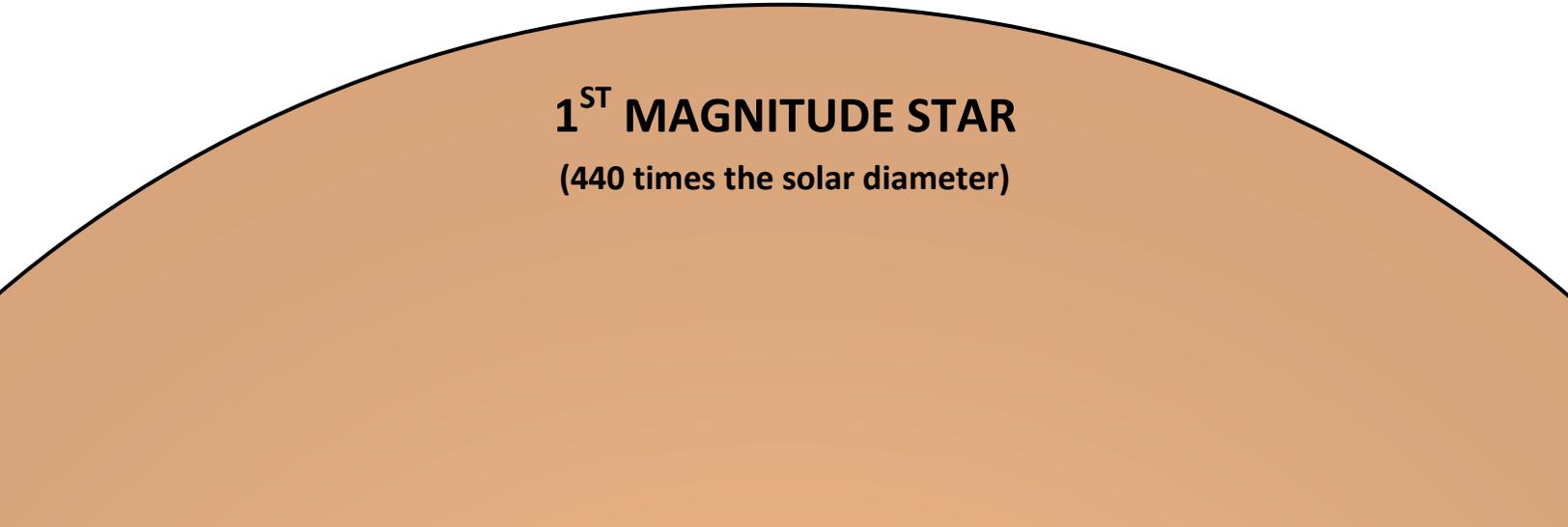

1<sup>ST</sup> MAGNITUDE STAR

(440 times the solar diameter)

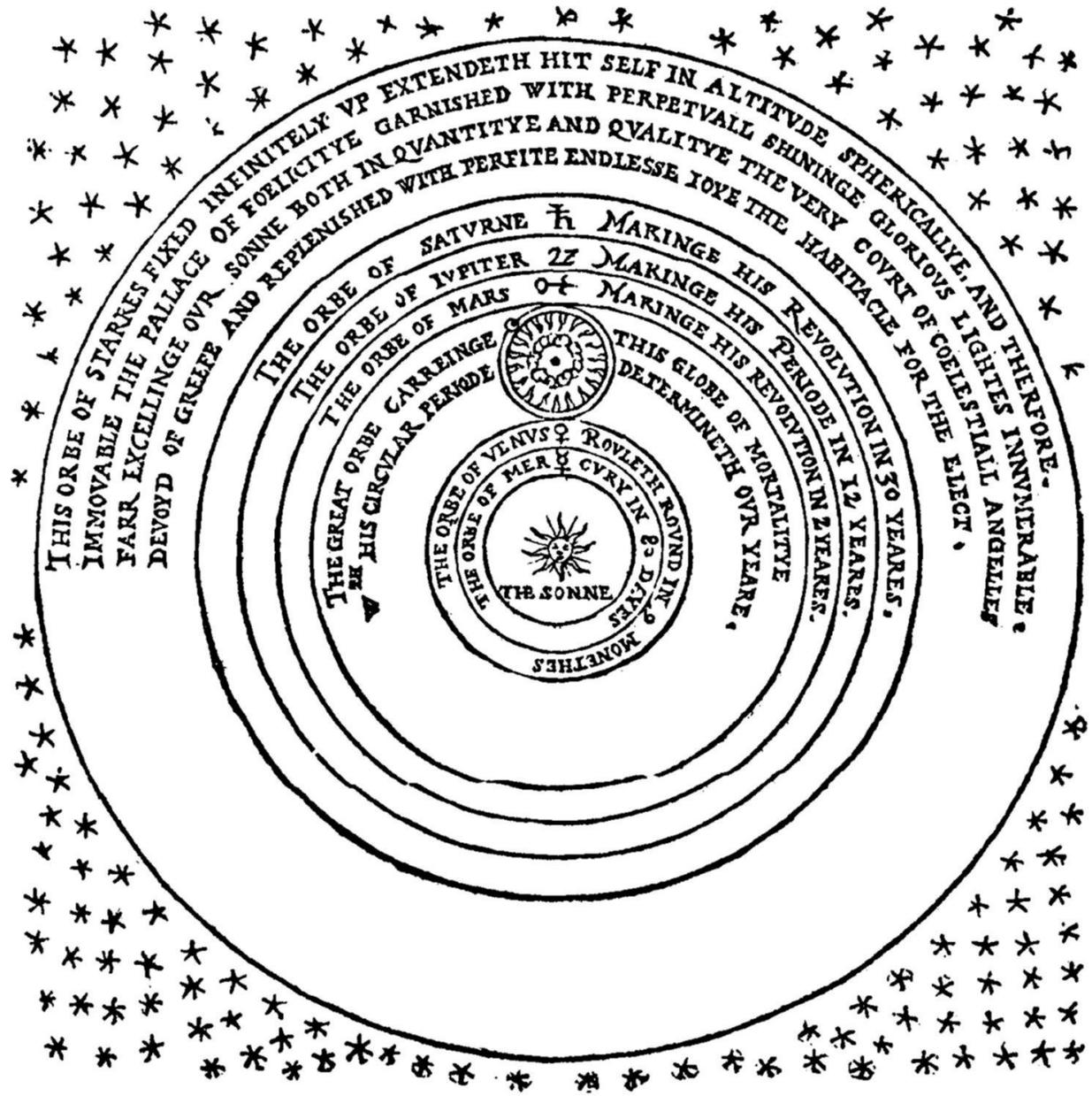

**FIGURE 3**

**Thomas Digges's representation of the Copernican theory from his 1576 "A perfit description of the Cælestiall Orbes". Note Digges's description of the starry heaven as "the palace of felicity garnished with perpetual shining glorious lights innumerable, far excelling our Sun both in quantity and quality, the very court of celestial angels, devoid of grief and replenished with perfect endless joy, the habitacle for the elect." Elsewhere he**

states that the starry heaven "may well be thought of us to be the glorious court of the great God, whose unsearchable works invisible, we partly by these his visible, conjecture; to whose infinite power and majesty, such an infinite place, surmounting all other both in quantity and quality, only is convenient." Digges was an accomplished astronomer and mathematician, who could do the calculations that indicate the necessity for giant stars in the Copernican theory. See Johnson & Larkey 1934.

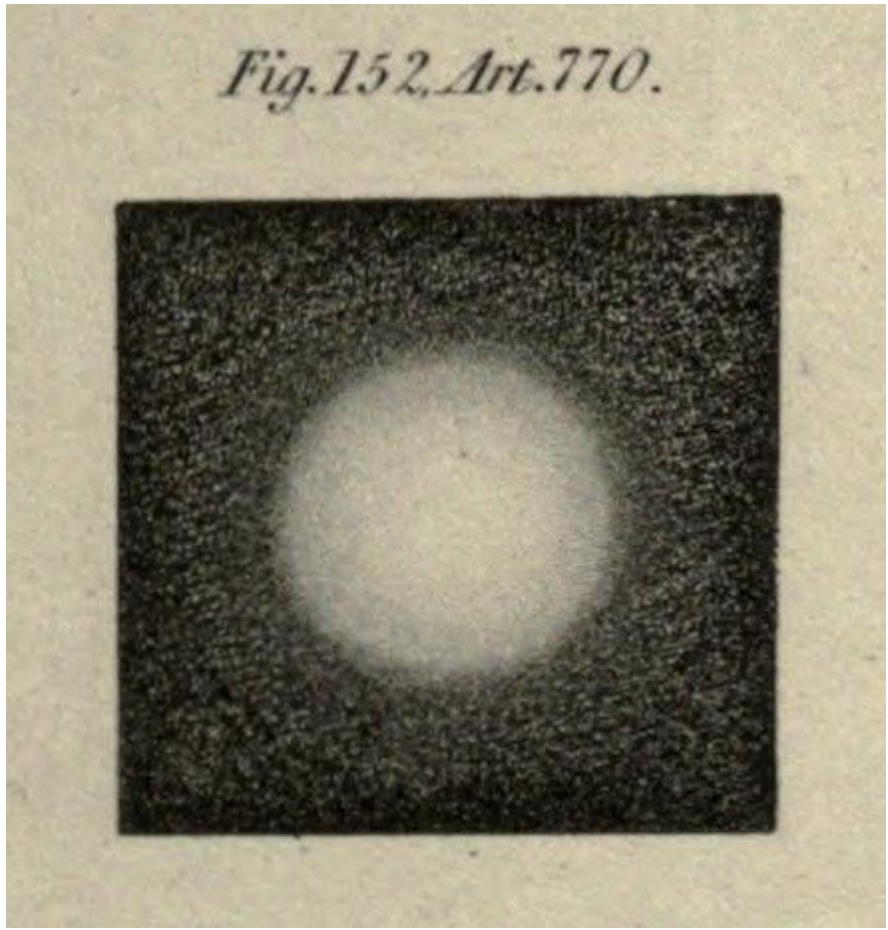

**FIGURE 4**

A star as seen through a small aperture telescope (see Herschel 1828, 491 & Plate 9). The spherical appearance is entirely spurious — an artifact of diffraction. However, early telescopic astronomers took such telescopic images to be the physical bodies of stars (Graney & Grayson 2011). The disk is smaller than what is seen with the naked eye, but as the telescope also increases the sensitivity to parallax, the final result is the same: Tycho, using naked-eye instruments, measured stars to have apparent diameters of roughly one minute of arc and calculated their distance and physical size based on a parallax of one minute; Riccioli, using a telescope, measured their diameters to be roughly one-sixth Tycho's diameters, and calculated distances and physical sizes based on one-sixth the parallax; in essence, each considered the threshold for detecting parallax to be one star diameter; thus both arrived at similar conclusions about the Copernican theory requiring absurdly large stars (Graney 2010b). As early as 1614 Simon Marius argued that disks of stars observed through the telescope undermined Copernicus and supported Tycho (Graney 2010a).